
\magnification\magstep 1
\vsize=8.5 true in
\hsize=6 true in
\voffset=0.5 true cm
\hoffset=1 true cm
\baselineskip=20pt

\bigskip
\bigskip
\bigskip
\vskip 5 truecm
\centerline{\bf SUM RULES AND BOSE-EINSTEIN CONDENSATION}

\vskip 1.5 truecm

\centerline{S. Stringari}
\bigskip
\centerline{{\it Dipartimento di Fisica, Universit\'a di Trento,
  I-38050 Povo, Italy}}
\vskip 2 truecm
{\it{\bf Abstract}. Various sum rules accounting
for the coupling between density and particle excitations and
emphasizing in an explicit way the role of the Bose-Einstein condensation
are discussed.
Important consequences on the fluctuations of the particle operator
as well as on the structure
of elementary excitations are reviewed. These include a recent
generalization of the Hohenberg-Mermin-Wagner theorem holding
at zero temperature.}

\bigskip
\bigskip

\par\noindent International Workshop on Bose-Einstein Condensation,
Levico (Trento), Italy, May 31-June 4, 1993

\vskip 1.0truecm

\bigskip
\bigskip

\vskip 1.5truecm

\par\noindent
\vfill\eject

{\parindent0pt{\bf 1.  INTRODUCTION}}

\bigskip

The sum rule approach has been extensively employed in the
literature in order to explore various dynamic features of
quantum many body systems from a microscopic point of view
(see for example [1] and references therein).
An important merit of the method is its explict emphasis
on the role
of conservation laws and of the symmetries of the problem. Furthermore
the explicit determination of sum rules is relatively easy and often requires
only a limited knowledge of the system.
Usually the sum rule approach is however employed without
giving special emphasis on
the possible occurrence of (spontaneously) broken symmetries.
For example the most famous $f$-sum rule [2] holding for a large class of
systems is not affected by the existence of an order parameter in
the system.

The purpose of this paper is to discuss a different class of sum rules
which are directly affected by the presence
of a broken symmetry.
These sum rules can be used to predict significant  properties of the system
which are the consequence of the existence of an order parameter.
In this work we will make explicit reference to Bose systems and to the
consequences of Bose-Einstein condensation (BEC).
Most of these results can be however generalized to discuss other systems
exhibiting spontaneously broken symmetries.

In sect. 2 we use a sum rule due to Bogoliubov in order to derive important
constraints on the fluctuations of the particle operator as well as
to obtain Goldstone-type bounds for the energy of elementary excitations.
In sect. 3 we explore the consequences of BEC on the long range behavior
of the half diagonal two-body density matrix and discuss
the coupling between the {\it density} and {\it particle} pictures
of elementary excitations in Bose superfluids.

\vfill\eject

\bigskip

\par\noindent
{\bf 2. THE BOGOLIUBOV SUM RULE.}

\bigskip

The Bogoliubov sum rule [3,4]
$$
\int^{+\infty}_{-\infty} {\cal A} _{a-a^{\dagger},\rho} (\omega)d\omega =
<[a_{\bf q} - a^{\dagger}_{-{\bf q}},\rho_{\bf q}]> = <a_0+a^{\dagger}_0>
= 2\sqrt{Nn_0}
\eqno(1)
$$
is the most remarkable among the sum rules depending explicitly on the order
parameter $n_0 = {1\over N} \mid<a_0>\mid^2$.
In eq.(1) ${\cal A}_{A^{\dagger},B}(\omega)$ is the spectral function
$$
{\cal A}_{A^{\dagger},B}(\omega) = {1\over Z} \sum_{m,n}
(e^{-\beta E_m} - e^{-\beta E_n})<m\mid A^{\dagger} \mid n> <n\mid B \mid m>
\delta(\omega - E_n - E_m)
\eqno(2)
$$
relative to the operators $A$, $B$ and we have made the choice
$A^{\dagger}=a_{\bf q} - a^{\dagger}_{-{\bf q}}$ and
$B=\rho_{\bf q}$
($\rho_q$, $a_{\bf q}$ and  $a^{\dagger}_{-{\bf q}}$ are the usual density,
particle annihilation and creation operators respectively
and we have taken $<a_0> = <a^{\dagger}_0> = \sqrt{Nn_0}$).
In deriving result (1)
we have used the completeness relation and the Bose commutation relations
for $a$ and $a^{\dagger}$. Despite its simplicity, result (1)
is at the basis of very important results characterizing  the macroscopic
behavior of Bose superfluids. Its richness is mainly due to the fact that
it exploits in a microscopic way the effects of the "phase"
operator, proportional
to the difference $a_{\bf q} - a^{\dagger}_{-{\bf q}}$.
Some important applications of result (1) are discussed below.

\bigskip
{\parindent0pt {\bf 2.1. The Hohenberg-Mermin-Wagner (HMW) theorem.}}
\bigskip
This theorem, of fundamental importance in condensed matter physics,
states that no long range order  can
exist ($n_0 = 0$ in the Bose case)
in a significant class of 1D and 2D systems at finite temperature [5-7].
The theorem is based on the Bogoliubov inequality [3,4]
$$
<\{A^{\dagger},A\}><[B^{\dagger},[H,B]]> \ge k_BT \mid<[A^{\dagger},B]>\mid^2
\eqno(3)
$$
for the fluctuations of the operator $A$ (we assume $<A> = <B> = 0$).
Inequality (3) can be  derived starting from the inequality [3,4]
$$
\chi_{A^{\dagger},A} <[B^{\dagger},[H,B]]> \ge \mid<[A^{\dagger},B]>\mid^2
\eqno(4)
$$
for the static response relative to the operator $A$
$$
\chi_{A^{\dagger},A}
= \int_{-\infty}^{+\infty}{\cal A}_{A^{\dagger},B}(\omega)
{1\over \omega} d\omega
$$
and the result $<\{A^{\dagger},A\}> \ge 2k_BT\chi_{A^{\dagger},A}$
following from the fluctuation dissipation theorem.

Choosing in the Bose case $A^{\dagger}=a_{\bf q}$ and $B=\rho_{\bf q}$
we obtain the inequality [5]
$$
n(q) \equiv <a^{\dagger}_{\bf q} a_{\bf q}> \ge n_0m {k_BT \over q^2}
- {1\over 2}
\eqno(5)
$$
for the momentum distribution $n(q)$ of the system.
In order to derive result (5) we have used the $f$-sum rule [2]
$$
\int \omega S(q,\omega)d\omega =
{1\over 2} <[\rho_{\bf q}^{\dagger},[H,\rho_{\bf q}]]> = N{q^2\over 2m}
\eqno(6)
$$
holding for Galilean invariant potentials. In eq.(6)
$S(q,\omega) = {\cal A}_{\rho^{\dagger},\rho}(\omega)/(1-exp(-\beta\omega))$
is the usual dynamic structure function. Inequality (5)
points out the occurrence of an important infrared $1/q^2$ divergency in the
momentum distribution which originates from
the thermal fluctuations of the phase of the condensate.
This behavior permits to prove the absence of BEC in 1D and 2D
systems at finite temperature. In fact, due to such a
divergency, the normalization condition
for the momentum distribution $\sum_{\bf q} n(q) = N$
cannot be satisfied in 1D and 2D, unless $n_0=0$ [5].

\bigskip

{\parindent0pt {\bf 2.2. Extension of the Hohenberg-Mermin-Wagner theorem
at T=0.}}
\bigskip

As clearly revealed by eqs.(3),(5) the HMW theorem
 does not apply at zero temperature. Actually there are important
examples of 2D systems obeying the HMW theorem at $T \ne 0$ and exhibiting
long range order in the ground state. The physical reason is that
quantum fluctuations have in general a weaker effect with
respect to thermal fluctuations in destroying the order parameter.

An extension of the Hohenberg-Mermin-Wagner theorem holding at zero
temperature has been recently
formulated by Pitaevskii and Stringari [8].
To this purpose one uses the uncertainty principle inequality
$$
<\{A^{\dagger},A\}><\{B^{\dagger},B\}>
\ge \mid<[A^{\dagger},B]>\mid^2
\eqno(7)
$$
rather than the Bogoliubov inequality (3). Both inequalities
provide rigorous constraints on the fluctuations of the operator $A$.
However, while the Bogoliubov
inequality is sensitive to themal fluctuations and becomes less and less
useful as $T\to 0$, the uncertainty principle is  particularly powerful
in the low temperature regime dominated by quantum fluctuations. By making
the choice $A^{\dagger}=a_{\bf q}$, $B=\rho_{\bf q}$ one obtains the following
non trivial inequality for the momentum distribution [8]:
$$
n(q) \ge {n_0 \over 4S(q)} - {1\over 2}
\eqno(8)
$$
of a Bose system where
$S(q) = {1\over N} <\rho_{\bf q}^{\dagger}\rho_{\bf q}>$ is the static
structure function. The low $q$ behavior of $S(q)$ is fixed,
at zero temperature, by the further bound
$$
S(q) \le \sqrt{\int \omega S(q,\omega) d\omega \int {1\over \omega} S(q,\omega)
d\omega}
\eqno(9)
$$
that can be easily calculated at small $q$. In fact in this limit the
inverse energy weighted sum rule (compressibility sum rule)
$ \int {1\over \omega} S(q,\omega) ={1\over 2}\chi(q) $ approaches
the compressibility parameter of the system ($lim_{q\to 0}
\chi(q) = 1/mc^2$),
while the energy weighted sum rule (f-sum rule) is given by eq.(6).
As a consequence $S(q) \le q/2mc$ at small $q$ and
the momentum distribution n(q) diverges at least as
$$
n(q) = n_0 {mc\over 2q} .
\eqno(10)
$$
It is worth pointing out that the $1/q$
law for the momentum
distribution, already known in the literature [9,10],
has been obtained here without any assumption on the nature
of the elementary excitations of the system (phonons in a neutral
Bose superfluid) and follows uniquely by the existence of BEC, the
validity of the f-sum rule and the finitness of the compressibility.
By imposing the proper normalization to the momentum distribution one can
then rule out the existence of BEC in 1D systems at zero temperature.
Note that the ideal Bose Gas does not violate the theorem since in this
case the compressibility sum rule $\chi(q)$ diverges as $1/q^2$.

Starting from the same inequality (7) it is possible to rule out the
existence of long range order also in other 1D systems at T=0 ( isotropic
antiferromagnets,
crystals) [8]. The theorem does not apply to 1D ferromagnets since in this case
the inverse energy weighted sum rule (magnetic susceptibility) diverges
as $1/q^2$.

Inequality (8) has been recently used [11] to rule out the existence of BEC
in the  bosonic representation of the electronic wave function
$$
\Psi_B({\bf r}_1, ..., {\bf r}_N) = exp( {i\over \nu}\sum_{i,j} \alpha_{i,j})
\Psi_F({\bf r}_1, ..., {\bf r}_N)
\eqno(11)
$$
proposed by Girvin  and MacDonald [12] for the fractional quantum Hall effect.
In eq.(11) $\Psi_F$ is the fermionic wave function
of electrons, $\nu = {1 \over 2k+1}$,
where $k$ is an integer, is the usual filling factor
and $\alpha_{i,j}$ is the angle between the vector connecting
particles $i$ and $j$  and an arbitrary fixed axis.
Using the Laughlin's expression [13] for the ground state wave function
$\Psi_F$ the authors of ref.[12]
were able to show that there is not Bose-Einstein
condensation in the bosonic wave function $\Psi_B$,
but only algebraic long range order. The same
result was obtained in ref.[14] starting directly from the
Chern-Simons-Landau-Ginzburg theory.
In ref.[11] the absence of long range order was proven starting directly
from the unceratinty principle inequality (8).
The result
follows from the fact that a charged liquid in an external magnetic field
is characterized by a suppression of density fluctuations resulting in a
quadratic law
$$
S(q) = {q^2 \over 2m\omega_c}
\eqno(12)
$$
for the static structure function ($\omega_c$ is the
cyclotronic frquency). This behaviour reflects the incompressibility
of the system and is consistent with the Kohn's theorem [15].
Inequality (8) implies a $1/q^2$
divergency for the momentum distribution of the bosonic wave function
of this 2D problem. This is incompatible with the normalization
of $n(q)$, fixed by the total number of electrons, unless the Bose
condensate relative to $\Psi_B$ identically vanishes.

\bigskip

{\parindent0pt {\bf 2.3. BEC and excitation energies.}}
\bigskip
The sum rule technique can be used to
to get useful constraints on the energy of elementary excitations.
In particular use of the Bogoliubov sum rule (1) makes it possible to
emphasize the role of the order parameter.

Given a pair of operators $A$ and $B$ one can derive the following rigorous
inequality, holding
at zero temperature, for the energy $\omega_0$ of the lowest state
excited by the operators
$A$ and $B$:
$$
\omega_0^2 \le {<[A^{\dagger},[H,A]]><[B^{\dagger},[H,B]]>
\over \mid<[A^{\dagger},B]>\mid^2}
\eqno(13)
$$
Bounds of the form (13) were first considered by Wagner [4].
Result (13) can be obtained using the inequality (holding at zero
temperature)
$$
\omega_0^2 \le {<[A^{\dagger},[H,A]]> \over \chi_{A^{\dagger},A}}
\eqno(14)
$$
where the r.h.s coincides with the ratio between
the energy weighted and the inverse energy weighted sum rules
relative to the operator $A$. Use of the Bogoliubov inequality (4) then
yields eq.(13).

Result (13) has the important merit of providing a rigorous upper bound
for $\omega_0$ in terms of quantities involving commutators.
This is an advantage, for example, with respect to the so called Feynman bound:
$$
\omega_0 \le {<[A^{\dagger},[H,A]]> \over <\{A^{\dagger},A\}>}
\eqno(15)
$$
involving the anticommutator $<\{A^{\dagger},A\}>$ in the denominator
 and hence requiring
the direct knowledge of the fluctuations of the operator $A$.
The occurrence
of the anticommutator makes it
difficult to exploit the low momentum regime of
elementary excitations. Viceversa eq.(13) can be directly employed
for this purpose.

Result (13) is particularly interesting when the expectation value of
 the commutator
$[A^{\dagger},B]$ is proportional to the order parameter of the
problem. This happens in a Bose superfluid with the choice
$A^{\dagger}=a_{\bf q} - a_{-{\bf q}}^{\dagger}$ and $B=\rho_{\bf q}$,
already considered in this work. One then finds
$$
\omega_0^2(q) \le {1\over n_0} {q^2 \over 2m} [{q^2 \over 2m} - \mu
+ NV(0) + \sum_{\bf p} V(p)(n({\bf p}+{\bf q}) + \overline{n}({\bf p}+{\bf
q}))]
\eqno(16)
$$
where we have carried out the commutators using the grand canonical hamiltonian
$H^{\prime} = H - \mu N$ ($\mu$ is the chemical potential) and taken a
central potential with Fourier transform $V(p)$. The quantitities $n(p)$ and
$\overline{n}(p)$ are defined by
$n(p) = <a^{\dagger}_{\bf p}a_{\bf p}>$ and
$\overline{n}(p) = {1\over2}<a^{\dagger}_{\bf p}a^{\dagger}_{-{\bf p}}
+a_{\bf p}a_{-{\bf p}}>$.

Result (16) is a rigorous inequality holding for any Bose system interacting
with central potentials. It has the form of a Goldstone theorem. In fact,
since its right hand side
behaves as $q^2$ when $q\to 0$,
it proves the existence of gapless excitations, provided
the order parameter $n_0$ is different from zero.

It is worth noting that in the limit of a dilute Bose gas
($\mu = NV(0), n(p) = \overline {n}(p) = N\delta_{{\bf p},0}, n_0=1$),
this upper bound coincides  with the Bogoliubov dispersion law
$$
\omega_B^2(q) = {q^2 \over 2m} [{q^2 \over 2m} + 2NV(q)]
\eqno(17)
$$
for any value of ${\bf q}$.
This result follows from the fact that the sum rules entering inequality (13)
are exhausted by a single excitation, multi-particle states playing a
negligeable role in a dilute gas. Of course in a strongly interacting system,
such as liquid $^4He$, multiparticle excitations are much more important
and the bound (16) turns out to be significantly higher than the lowest
excitation energy $\omega_0(q)$.

It is useful to compare the Goldstone type inequality (16)
with another, also rigorous, upper bound,
still derivable from eq.(13), by choosing
$$
A =
{\bf q \cdot j_{\bf q}}
= [H,\rho_{\bf q}] =
{1\over 2m} \sum_{\bf k} (q^2+2{\bf q\cdot k})
a^{\dagger}_{{\bf k}+{\bf q}}a_{\bf k}
\eqno(18)
$$
and $B=\rho_{\bf q}$. In eq.(18) ${\bf j}_{\bf q}$ is the usual
current operator.
The resulting bound then
coincides with the ratio between the cubic energy weighted and the energy
weighted sum rules relative to the density operator $\rho_{\bf q}$ [16,17]
$$
\omega_0^2(q) \le {<[[\rho_{-{\bf q}},H],[H,[H,\rho_{\bf q}]]]> \over
<[\rho_{-{\bf q}},[H,\rho_{\bf q}]]>} \, .
\eqno(19)
$$
The explicit calculation of the triple commutator yields (we choose ${\bf q}$
along the $z$-axis) [16,17]
$$
<[[\rho_{-{\bf q}},H],[H,[H,\rho_{\bf q}]]]>
$$
$$
= N{q^2 \over m}
[({q^2\over 2m})^2 + {2q^2\over m}<E_K> + {\rho\over m}
\int d{\bf s} (1-cosqz)g(s)\nabla_z^2V(s)]
\eqno(20)
$$
where $<E_K>$ is the ground state kinetic energy and $g(s)$ is the
pair correlation function.
The bound (19) then becomes:
$$
\omega_0^2(q) \le ({q^2\over 2m})^2 + {2q^2\over m}<E_K> + {\rho\over m}
\int d{\bf s} (1-cosqz)g(s)\nabla_z^2V(s)
\eqno(21)
$$
and exhibits a quadratic behaviour in $q$ as $q\to 0$.
It is worth noting that in the limit of a dilute Bose gas
($<E_K>=0, g(s)=1$)
both bounds (21) and (16)
coincide
with the Bogoliubov dispersion (17).

The fact that it is possible to prove, via eq.(21),
the existence of gapless excitations
without assuming the existence of a broken symmetry is
directly connected (see eq.(19)) with the conservation of the total current
${\bf j}_{{\bf q}=0}$,
holding in translationally invariant systems. This discussion  also
suggests that Goldstone type inequalities of the form (16) are
particularly useful in sytems where the current is ${\bf not}$ conserved
and where consequently only the existence of a spontaneously broken symmetry
can ensure in a simple way the occurrence of gapless excitations.
This is the case, for example, of spin excitations in magnetic systems
(spin current is not conserved) or Bose systems with random external
potentials.

Let us discuss for example the effects of an external
potential of the form (we take $U_{-{\bf k}} = U^*_{\bf k}$)
$$
V_{ext} = \sum_{{\bf k} \ne 0} U_{\bf k}\rho_{\bf k}
\eqno(22)
$$
on the sum rules discussed above. A first important result is that
the Goldstone type upper bound
(16) is not directly affected by the external field because of the exact
commutation property
$$
[a_{\bf q} - a^{\dagger}_{-{\bf q}},[V_{ext},a^{\dagger}_{\bf q}
- a_{-{\bf q}}]] = 0 \, .
\eqno(23)
$$
Viceversa the cubic energy weighted sum rule for the density operator
$\rho_{\bf q}$  gets an extra contribution
from the external force given by
$$
<[[\rho_{-{\bf q}},H],[V_{ext},[H,\rho_{\bf q}]]]>
= {q^2 \over m^2}
<-\sum_{\bf k}k^2_z U_{\bf k}\rho_{\bf k}>
\eqno(24)
$$
At small $q$ the new term provides the leading contribution to the triple
commutator (20). Since the quantity
$<[\rho_{-{\bf q}},[H,\rho_{\bf q}]]>$ is not changed by the external force,
being still given by the $f$-sum rule (6), the upper bound (19)
no longer vanishes with $q$.
This different behaviour is due to the fact
that the current ${\bf j}_{{\bf q}=0}$ is not conserved
in the presence of the external field (22).
This  result also reveals that the relationship
${\bf j_q} = {\bf q}{\sqrt{Nn_0}
\over 2m} (a^{\dagger}_{\bf q} - a_{-{\bf q}})$
between the current and the gradient of the phase operator, holding
in a translationally invariant dilute Bose gas, is no longer
valid in the presence
of an external potential and gives rise to
a normal (non superfluid) component of the density of the system
at zero temperatrure. This is discussed, for example,
in the Huang's contribution to this workshop.

\bigskip

{\parindent0pt {\bf 3. THE HALF DIAGONAL TWO-BODY DENSITY MATRIX.}}
\bigskip

 In this section we discuss another sum rule, also sensitive to
the presence of Bose Einstein condensation, given by [18]
$$
\int _{-\infty}^{+\infty} {1 \over 1- e^{-\beta\omega}}
{\cal A}_{a+a^{\dagger},\rho}(\omega) d\omega
= <(a_{{\bf q}}
+a_{-{\bf q}}^{\dagger})\rho_{{\bf q}}>
\eqno(25)
$$
where  ${\cal A}$ is the spectral function already introduced in sect.1
with the choice $A^{\dagger}=a_{{\bf q}}+a^{\dagger}_{-{\bf q}}$
and $B=\rho_{\bf q}$.

Differently from eq.(1), accounting for the commutation relation between the
density and the phase operators, this sum rule
cannot be expressed in terms of a commutator.
Physically it accounts for the coupling between the density of
the system and the modulus
of the condensate, proportional to
$a_{\bf q}+a^{\dagger}_{-{\bf q}}$.
An interesting feature of this sum rule is
that it fixes the long range behavior of the half-diagonal two-body
density matrix
$$
\rho^{(2)}({\bf r}_1, {\bf r}_2;{\bf r}_1^{\prime}, {\bf r}_2)
= <\Psi^{\dagger}({\bf r}_1)
\Psi^{\dagger}({\bf r}_2)\Psi({\bf r}_1^{\prime})\Psi({\bf r}_2)>
\eqno(26)
$$
The occurrence of BEC is in fact not only relevant for the long range behavior
of the one-body density matrix
$$
\rho^{(1)}({\bf r},{\bf r}^{\prime}) =
<\Psi^{\dagger}({\bf r})\Psi({\bf r}^{\prime})>
\eqno(27)
$$
fixed by the law
$$
\lim_{{\bf r}^{\prime} \to \infty}\rho^{(1)}({\bf r},{\bf r}^{\prime})
= \rho n_0 \, ,
\eqno(28)
$$
but also for the one of the two-body matrix (26).
The LRO in the 2-body density matrix (26) is naturally defined by
the equation [19]
$$
\lim_{{\bf r}^{\prime}_1 \to \infty}\rho^{(2)}({\bf r}_1,
{\bf r}_2;{\bf r}_1^{\prime}, {\bf r}_2)
= n_0\rho^2 (1 + F_1(\mid{\bf r}_1-{\bf r}_2\mid))
\eqno(29)
$$
and is characterized by the condensate fraction $n_0$ as
well as by the function $F_1(r)$. The properties of this
function have been recently investigated in ref.[18,19].
The link between the sum rule (25) and the  long range behavior of
$\rho^{(2)}$ is fixed by the relation
$$
<\rho_{{\bf q}}(a_{{\bf q}}
+a_{-{\bf q}}^{\dagger})>  = \sqrt{Nn_0}(1+2F_1(q))
\eqno(30)
$$
where $F_1(q)$ is the Fourier transform of $F_1(r)$. At low $q$
the behavior of this function is fixed by the density
dependence of the condensate according to law [18]
$$
\lim_{q\to 0}{1 + 2F_1(q) \over q} = {1\over 2n_0mc}{\partial (n_0 \rho)
\over \partial \rho} \, .
\eqno(31)
$$

The occurrence of LRO in the two-body density matrix implies the existence
of a non trivial relation for the chemical potential that can be
obtained starting from the
 following expression holding in systems exhibiting Bose-Einstein
condensation:
$$
\mu = E(N) - E(N-1) = <H> - {<a^{\dagger}_0 H a_0> \over <a^{\dagger}_0 H a_0>}
\, .
\eqno (32)
$$
Result (32) for the chemical potential
follows from the property that the  ${\bf p}=0$ state
in a Bose superfluid plays the role of a {\it reservoir} and that
consenquently adding (or destroying) a particle in this state
yields the equilibrium state relative to the $N+1$ ($N-1$) system.
Starting from eq.(32) one easily derives the equation:
$$
\mu = -{<a^{\dagger}_0 [H,a_0]> \over <a^{\dagger}_0 a_0>} =
 {1\over \sqrt{Nn_0}} \sum _{\bf p} <a^{\dagger}_{-{\bf p}}\rho_{\bf p}> V(p)
\/ .
\eqno (33)
$$
Using results (1) and (30) one then finds the
exact non trivial relationship [18]
$$
\mu = \rho\int d{\bf r} V(r) (1+F_1(r))
\eqno (34)
$$
relating the LRO function $F_1(r)$ to the chemical potential
of the system.
Result (34) holds for any Bose system exhibiting condensation and interacting
with central potentials.

The LRO function $F_1(q)$ plays a crucial role  in the coupling
between the {\it density}
$$
\mid F> = {1\over \sqrt{NS(q)}}\rho_{\bf q}\mid 0>
\eqno(35)
$$
and {\it particle}
$$
\mid P> = {1\over \sqrt{n(q)}}a_{-{\bf q}}\mid 0>
\eqno(36)
$$
states that provide natural approximations to the
elementary excitations of a Bose system.
Result (35) coincides with the most famous Bijl-Feynman ansatz [20].
 Both the {\it density} and {\it particle} pictures coincide
with the exact eigenstates in the
limit of a dilute Bose gas. For a strongly correlated liquid
they provide only an approximate
description. The coupling between
the two states is different from zero because of occurrence
of the Bose-Einstein condensation
and is given by [1]:
$$
<F\mid P> = \sqrt{{n_0 \over S(q)n(q)}} F_1(q)
\eqno (37)
$$
The coupling turns out to be complete when $q\to0$
[in fact in this limit one has $S(q)= q/2mc, n(q)= n_0mc/2q$ and
$F_1(q)=-{1\over2}$] showing in an explict way
that in the hydrodynamic limit the {\it density} and {\it particle} picture
of elementary excitations of a Bose condensed system coincide.
The coupling between the {\it density} and {\it particle} pictures
is at the basis of fundamental properties exhibited by Bose superfluids
(for a recent exhaustive discussion see ref.[21]).

It is finally interesting to compare the average excitations
energies of the states (35) and (36). Both energies provide a rigorous
upper bound to the energy of the lowest excited state of the system.
The energy of the Feynman state (35) is given
by the most famous law [20]
$$
\epsilon_F(q)= {<F\mid H\mid F>\over<F\mid F>} = {q^2\over 2mS(q)}
\eqno(38)
$$
and is expressed in terms of static structure function.
It is well known that in liquid $^4$He eq.(38) provides the exact
dispersion in the low $q$ phonon regime, while it gives only a
poor description at higher momenta.

The energy of the particle state (36) takes instead the form [1]
$$
\epsilon_P(q) = \mu-{q^2 \over 2m}  - {W(q) \over n(q)}
\eqno (39)
$$
where $\mu$ is the chemical potential and the quantity $W(q)$ is given by
$$
W(q) = \int d{\bf r}_1 d{\bf r}_1^{\prime}
\rho^{(2)}({\bf r}_1,0;{\bf r}^{\prime}_1,0)
e^{-i{\bf q}({\bf r}^{\prime}_1 - {\bf r}_1)}V(r_1)
\eqno(40)
$$
with $\rho^{(2)}({\bf r}_1,0;{\bf r}^{\prime}_1,0)$ defined
in eq.(26).

It is also useful to note that, due to relation (33-34) for the chemical
potential, the particle energy (39)
has no gap at $q=0$ and vanishes linearly with
$q$ [1].

It is possible
to obtain an explicit expression for the average value
of the particle energy (39) in momentum space:
$$
\overline {\epsilon}_P = {\sum_{\bf q} n(q)\epsilon_P(q) \over \sum_{\bf q}
 n(q)} \, .
\eqno(41)
$$
This average is sensitive to the values of $\epsilon_P(q)$ in the
interval of momenta where the quantity $q^2n(q)$ has a significant
weight. In superfluid $^4He$ this corresponds to the range
$q=1-3A^{-1}$ including the maxon and roton region.
Using the operator identity [22]
$-\sum_{\bf p}a^{\dagger}_{\bf p}[H,a_{\bf p}]=E_K+2V$ the average (41)
takes the form
$$
\overline {\epsilon}_P = \mu -{1\over N}(<E_K> +2<V>)
\eqno(42)
$$
where $<E_K>$ and $<V>$ are the kinetic energy and the
potential energy   relative to the ground state.
At zero pressure, where $\mu={1\over N}(<E_K>+<V>)$,
eq.(42) yields $\overline {\epsilon}_P = -<V>/N = 21\sim 22K$ in superfluid
$^4$He. It is instructive to compare the above value with the corresponding
average of the Feynman energy:
$$
\overline {\epsilon}_F = {\sum_{\bf q} n(q)\epsilon_F(q) \over \sum_{\bf q}
 n(q)} \, .
\eqno(43)
$$
Using microscopic estimates  for $S(q)$ and $n(q)$ we find
$\overline {\epsilon}_F =24-25K$ a value
rather close to $\overline {\epsilon}_P$.

The fact that the energies of the {\it density} and {\it particle} states
(35-36) turn out to be comparable in the most
relevant region $q=1-3A^{-1}$
reveals the importance of a careful microscopic
investigation of
the coupling (induced by BEC) between the two pictures
in order to get a better understanding of the nature
of elementary excitations in Bose superfluids.

\bigskip
\bigskip

\par\noindent REFERENCES

\bigskip

\item{1.} S. Stringari, Phys. Rev. B {\bf 46}, 2974 (1992);

\item{2.} D. Pines and Ph. Nozieres, {\it The Theory of Quantum Liquids}
(Benjamin, New York 1966),Vol.I; Ph. Nozieres and D. Pines {\it
The Theory of Quantum Liquids} (Addison-Wesley, 1990),Vol.II;

\item{3.} N.N. Bogoliubov, Phys. Abh. SU {\bf 6}, 1 (1962);

\item{4.} H. Wagner, Z. Physik {\bf 195}, 273 (1966);

\item{5.} P.C. Hohenberg, Phys. Rev. {\bf 158}, 383 (1967);

\item{6.} N.D. Mermin and H. Wagner Phys.Rev.Lett. {\bf 17}, 1133 (1966);

\item{7.} N.D. Mermin, Phys. Rev. {\bf 176}, 250 (1968);

\item{8.} L. Pitaevskii and S. Stringari, J. Low Temp. Phys. {\bf 85}, (1991);

\item{9.} T. Gavoret and Ph. Nozieres, Ann. Phys. (NY) {\bf 28}, 349 (1964);

\item{10.} L. Reatto and G.V. Chester, Phys. Rev. {\bf 155}, 88 (1967);

\item{11.} L. Pitaevskii and S. Stringari, Phys. Rev. B {\bf 47}, 10915,
(1993);

\item{12.} S. Girvin and A. MacDonald, Phys. Rev. Lett. {\bf 58}, 1252 (1987);

\item{13.} R.B. Laughlin, Phys.Rev.Lett. {\bf 50}, 1395 (1983);

\item{14.} S.C. Zhang, Int. J. Mod. Phys. {\bf 6}, 25 (1992);

\item{15.} W. Kohn, Phys.Rev. {\bf 123}, 1242 (1961);

\item{16.} R.D. Puff, Phys. Rev. A {\bf 137}, 406 (1965);

\item{17.} D. Pines and C.-W. Woo, Phys. Rev. Lett. {\bf 24}, 1044 (1970);

\item{18.} S. Stringari. J. Low Temp. Phys. {\bf 84}, 279 (1991);

\item{19.} M.L. Ristig and J. Clark, Phys. Rev. B {\bf 40}, 4355 (1989);

\item{20.} A. Bijl, Physica {\bf 8}, 655 (1940);
R.P. Feynman, in {\it Progress in Low Temperature Physics},
vol.1, ed. by C.J. Gorter (North Holland, Amsterdam, 1955), Ch.2;

\item{21.} A. Griffin, {\it Excitations in a Bose-Condensed Liquid}
(Cambridge University Press, 1993);

\item{22.} A.L. Fetter and J.D. Walecka, {\it Quantum Theory of Many-Particle
Systems} (McGraw-Hill, New York, 1971).

\bye